\documentclass[twoside]{article}
\usepackage{fleqn,espcrc2,epsfig}

\newcommand{\AmS}{{\protect\the\textfont2
  A\kern-.1667em\lower.5ex\hbox{M}\kern-.125emS}}
\title{Heat  and Charge Transport Properties of   MgB$_{\lowercase{2}}$ }
\author{Matthias Schneider\address[ITTP]
{Institut f\"ur Tieftemperaturphysik, Technische Universit\"at
Dresden, D-01062 Dresden, Germany}
\thanks{E-mail: schneid@physik.phy.tu-dresden.de},
Dieter Lipp\addressmark,
Alexander Gladun\addressmark,
Peter Zahn\address[ITP]{Institut f\"ur Theoretische Physik, Technische Universit\"at
 Dresden, D-01062 Dresden, Germany},
Axel Handstein\address[IFW]{Institut f\"ur Festk\"orper- und Werkstofforschung e. V.,
D-01171 Dresden, Postfach 270116, Germany},
G\"unter Fuchs\addressmark,
Stefan-Ludwig Drechsler\addressmark, Manuel Richter\addressmark ,
Karl-Hartmut M\"uller\addressmark ,
and Helge Rosner\address[UC]{Department of Physics, University of California,
Davis, CA 95616, USA}}

\begin{document}

\begin{abstract}
\footnotesize{
A polycrystalline sample of the MgB$_{2}$ superconductor was investigated by
measurements of the electrical resistivity, the thermopower and the thermal
conductivity
in the temperature range between 1.8K and 300K in zero magnetic field. The
electrical resistivity shows a  superconducting  transition at $T_c=38.7$K
and, similarly to borocarbides, a $T^{2.4}$ behaviour up to 200K.
The
electron diffusion thermopower  and its bandstructure-derived value
indicate the dominant hole character of the
charge carriers. The total  thermopower can be explained by the  diffusion term
renormalized by a significant electron-phonon interaction
and a phonon drag term. In the thermal conductivity,
for decreasing temperature,
 a significant decrease below
$T_c$ is observed resulting in a $T^{3}$ behaviour below 7K. The
reduced Lorenz number exhibits values smaller than 1 and a characteristic
minimum
which resembles the behaviour of non-magnetic borocarbides. }
\vspace{1pc}

\end{abstract}

\maketitle

\section{INTRODUCTION}
After the surprising discovery of superconductivity up to about 40K in MgB$_{2}$
\cite{Nagamatsu}  extensive investigations of its physical properties
have been performed.  Special interest is focused on the
electronic structure and in particular, on the type of the charge carriers, and
their relationship to
the superconducting pairing mechanism. Numerous studies
are devoted to thermodynamic properties such as
the specific heat \cite{budko1,Kremer,W"alti,Junod} and the
upper critical field
\cite{Finnemore,Mueller,Handstein,Fuchs,shulga01}.

However, there are
less reports published on the transport properties of MgB$_{2}$
and only few on the heat transport.
Results of previous
investigations of the electrical resistivity $\rho$  differ not only
in the
residual resistivity but also in the temperature dependence
 \cite{Finnemore,Canfield_neu,Jung,Gasparov}.
First measurements of the thermopower $S$ \cite{Lorenz,Liu,Choi} and
the thermal conductivity $\kappa$  \cite{Michor} show a significant
non-linearity in $S(T)$ in the temperature range close to room temperature and
rather high values for the Lorenz number derived from the reported data of
$\kappa (T)$ \mbox{and $\rho (T)$.}
Such measurements are of general interest since they provide
additional insight into the electronic structure
and the electron-phonon interaction.
In the present paper, zero magnetic field measurements of thermal and charge
transport properties
of MgB$_{2}$ are reported.
Since at present the pairing mechanism has not been settled yet the
comparison with
related superconductors might be helpful
to elucidate further details of the superconductivity in MgB$_2$. In this
context
 also similarities and differences with the behaviour of well studied non-magnetic
borocarbides
will be discussed.

\section{EXPERIMENTAL DETAILS}

The experiments were performed on a polycrystalline sample of MgB$_{2}$
of about \mbox{$5\times 1.2\times1.2\mathrm{mm}^3$.} It was cut from a pellet
which was prepared by a conventional solid state reaction as described
\mbox{elsewhere \cite{Mueller}.} The x-ray diffraction pattern of powder ground from this sample batch have
shown that the material is single phased.

To investigate the thermopower $S$, two copper wires
were fixed to the sample by an electrically conducting epoxy resin.
The temperature gradient
along the sample of about $1\%$ of the temperature
 was generated by a small strain gauge heater. The temperature
differences between the copper wires and between sample and cold copper plate
were measured by two AuFe-Chromel thermocouples, the absolute temperature
was detected by
a Germanium and at higher values by a Platinum thermometer.

The thermal conductivity $\kappa$ was measured by the standard steady-state
method.
For a first measurement up to 100K
the thermocouples were fixed directly to the sample
by a low temperature varnish. The same varnish was used to connect the sample
with the cold copper plate.
A second measurement, with a better contact by electrically conducting
epoxy resin
\mbox{Eccobond 56C,}
was performed together with the investigation \mbox{of $S$.}
No significant deviations between both results were found.

The electrical resistance was measured by the usual four probe method.
Unfortunately, the attempt to fix additional current contacts to the sample
prepared for the thermal conductivity measurement
failed. Therefore, the electrical resistivity $\rho$ was determined in a separate
run, resulting in higher errors for the reduced Lorenz number because of the higher
uncertainty in the distance between the voltage leads.

\section{RESULTS AND DISCUSSION}

As shown in \mbox{Fig. 1,} the resistivity of the investigated sample
decreases from room
temperature down to 40K from a value of 38.2$\mu \Omega$cm to
7.1$\mu \Omega$cm, {\it i. e.\/} the
resistivity ratio RRR amounts to 5.4. According to the uncertainties of the
cross-section of the sample and the
distance between the voltage contacts,
the error of the $\rho$ values is about 20\% ; the uncertainty
of RRR is much smaller.
M\"uller {\it et al.} reported RRR=4.5
for another sample cut from the same pellet \cite{Mueller}.
A sharp superconducting transition is found at 38.7K
 (midpoint value of the normal-state
resistivity).

\begin{figure}[bt]
\begin{center}
\epsfig{file=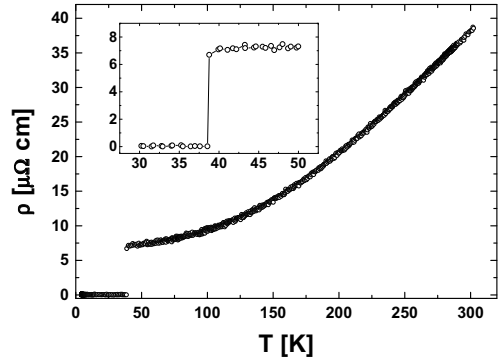,width=8.0cm}
\caption{\footnotesize{Temperature dependence of the electrical
resistivity $\rho$ of MgB$_{2}$.  The inset shows the range near $T_{c}$. }}
\label{fig1}
\end{center}
\end{figure}
To analyse the temperature dependence of $\rho(T)$, the normal state
data below 200K
can be fitted to the expression
\begin{equation}
\rho(T)=\rho_0+aT^b,
\label{rho}
\end{equation}
rather than to a Bloch-Gr\"uneisen
formula as proposed by Gasparov \mbox{{\it et al.\/} \cite{Gasparov}.}
\mbox{Fig.\ 2} shows the results in the
temperature range between $T_c$ and 200K.
The parameter values obtained from this fit
are $\rho_0=6.8\mathrm{\mu\Omega cm}$,
$a=3.3\times 10^{-5}\mathrm{\mu\Omega cm/K}^b$ and $b=2.4$.
The obtained value of the exponent is in
between the reported results $b=2$ \cite{Jung} and $b=3$ \cite{Finnemore} and in
good agreement with $b=2.6$ for a dense MgB$_{2}$ wire \cite{Canfield_neu}.
\begin{figure}[bt]
\begin{center}
\epsfig{file=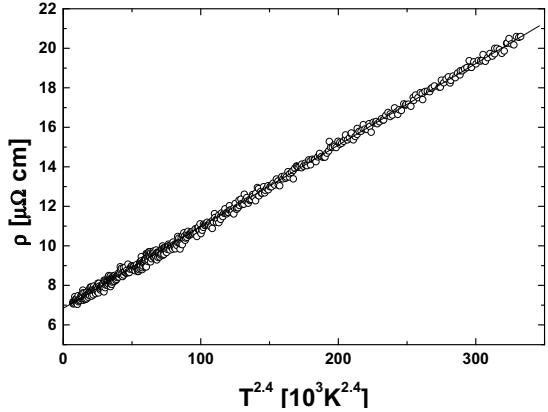,width=8.0cm}
\caption{\footnotesize{Resistivity $\rho$ of MgB$_{2}$ plotted as a function
of $T^{2.4}$. The plotted range corresponds to \mbox{$T_c<T<200$K.}  }}
\label{fig2}
\end{center}
\end{figure}
It is noteworthy that a similar behaviour was also found for other
superconducting compounds.
Rathnayaka {\it et al.\/} \cite{Canfield}  reported exponents $b$ of 2.2 and
2.0 for YNi$_2$B$_2$C and LuNi$_2$B$_2$C single crystals, respectively.
In the temperature range \mbox{$200$K$<T<300$K}
the curvature of $\rho (T)$
of MgB$_2$ decreases with increasing temperature 
and seems to follow the Bloch-Gr\"uneisen formula.

The thermopower of MgB$_{2}$ is shown in \mbox{Fig. 3.} The room temperature
 value of
8.7$\mu $V/K is in good agreement with the data of Lorenz
\mbox{{\it et al.\/}
\cite{Lorenz}} who reported a value of about 8.3$\mu $V/K for a
MgB$_{2}$ sample with a RRR of \mbox{about 3.}
Furthermore, the value for the slope of 0.036$\mu $V/K$ ^2$, fitted
to the measured data in the range
\mbox{40K$<T<$160K}, is close to the reported \mbox{d$S/$d$T=0.042\mu $V/K$ ^2$} below 160K \cite{Lorenz}.

\begin{figure}[hbt]
\begin{center}
\epsfig{file=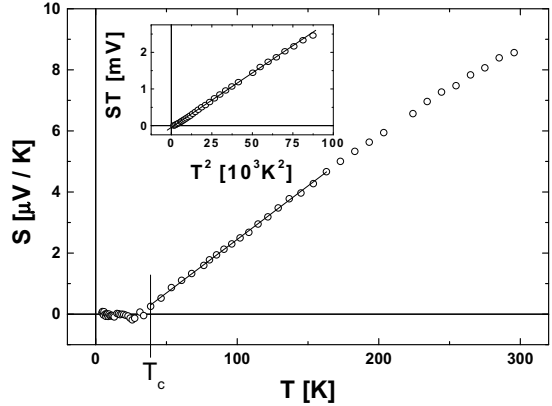,width=8.0cm}
\caption{\footnotesize{Thermopower $S$ of MgB$_{2}$ in the temperature range
between 4K and 300K. The straight line represents a linear fit to the data
in the range \mbox{40K$<T<$160K.}
The inset shows the product of thermopower and temperature as a function
of $T^2$ in the whole measured range. The straight line in the inset
represents a fit according \mbox{to Eq.\ \ref{tk}}
to the data in the range \mbox{70K$<T<270$K.} }}
\label{fig3}
\end{center}
\end{figure}

A quite smooth
behaviour of the thermopower with a small jump of about 0.3$\mu $V/K was found
at the crossover from
the superconducting to the normal state.
Lorenz {\it et al.\/} reported a higher jump of about 0.7$\mu $V/K \cite{Lorenz}.
Further investigations of
 high-quality very pure samples are required
to clarify this
issue.

The data can be
described by the expression
\begin{equation}
S(T)=\frac{A}{T} +BT,
\label{tk}
\end{equation}
where $A/T$ is the phonon drag term and $BT$ the electron diffusion term.
In the range \mbox{70K$<T<$270K},
for $B$ a value of $0.031\mu $V/K$^2$ is observed. The positive sign of $B$
is an indication of the hole character of the charge \mbox{carriers
\cite{Barnard,Blatt,Hirsch}.}
For A, the fit yields a value of about \mbox{-60$\mu $V.}
For temperatures below 70K a systematic deviation from the behaviour
according to Eq.\ \ref{tk} is found since
the expression $A/T$ is valid only at
the high temperature side of the phonon drag peak.
Above 270K, a sublinear behaviour of $ST${ \it vs.\/ }$T^2$ is observed
as reported by Lorenz {\it et al.\/} \cite{Lorenz}. The different
magnitude of this deviation might be attributed to differences in the samples.

To get some microscopic insight into the magnitude of the
second term of Eq.\ \ref{tk},
we have also performed a theoretical calculation of the electronic part
using LDA (local density approximation) bandstructure calculations
and employing the well-known Mott formula, renormalized
by the electron-phonon (el-ph) interaction coupling constant $\lambda_{el-ph}$
\cite{kaiser}:
\begin{equation}
S_{el}=\frac{\pi^2 k^2_B T}{3e}
\left[ \frac {\partial \ln \sigma (\varepsilon )}
{\partial \varepsilon}\right]_{\varepsilon =E_F}(1+\lambda_{el-ph}(T)).
\end{equation}
Here for the sake of simplicity only the energy dependence of the
conductivity $\sigma (\varepsilon) $ in the relaxation time approximation
 has been taken into
account ignoring a possible energy dependence of the scattering rates.
 Thus we obtain a value of about
\mbox{2.8 $\mu$V/K$\times (1+\lambda_{el-ph})$} at room
temperature. First of all the correct sign (which corresponds to a dominant
hole contribution) should be noted. Then,
adopting a strong el-ph interaction $\lambda_{el-ph} \sim 2$
we would arrive approximately at the experimental results \mbox{$\sim 8 \mu$V/K.}
This is in qualitative accord with the intermediate to strong coupling
scenario proposed
in Ref.\ \cite{shulga01}.
However, a more detailed investigation of each Fermi surface sheet and
of the coupling to various phonon (boson) modes
are required to extract quantitatively
 the strength of the el-ph interaction in a more reliable manner.
 In this context a possible relation of the (decreasing) temperature
 dependence of $\lambda_{el-ph} (T)$ at high temperatures to the
 observed
deviation from the behaviour according to Eq.\ \ref{tk} above 270K
is worth to be studied in more detail.

For YNi$_2$B$_2$C and LuNi$_2$B$_2$C the electron diffusion term
is smaller and negative:
For YNi$_2$B$_2$C, a 
value of about \mbox{$-0.007\mu $V/K$^2$
\cite{Fisher}}
is reported.
Furthermore, much higher negative phonon drag contributions in borocarbides
have been found
resulting in values for
$A$ between \mbox{$-450$} and \mbox{$-550\mu $V  \cite{Fisher}.}
Thus, the phonon drag contribution is less pronounced in MgB$_2$ than in
YNi$_2$B$_2$C.

The results of the thermal conductivity measurements
are presented in \mbox{Fig. 4. }
The data taken in two separate runs
as described above
lie on top up of each other.
Nevertheless, the error of the
absolute value is about 20\%,
mainly caused by
the uncertainties of
cross-section and distance
between thermometers.

The measured values of $\kappa$
at 300K
are about $20\%$ smaller than those reported by Bauer \mbox{{\it et al.\/} \cite{Michor}}
resulting from
uncertainties in the measurements and
possible differences in the samples.
The positive slope of $\kappa(T)$ in the whole investigated temperature range
indicates the limitation of the heat conductivity by crystal defects as in pure normal
metals $\kappa$ exhibits a maximum at lower temperatures and then decreases 
to a constant value 
with rising temperature.

\begin{figure}[hbt]
\begin{center}
\epsfig{file=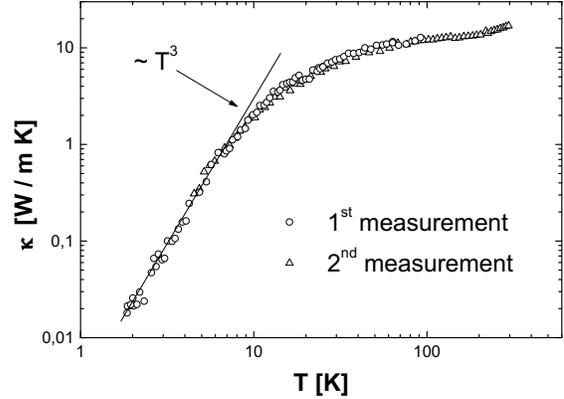,width=8.0cm}
\caption{\footnotesize{Thermal conductivity $\kappa$
of MgB$_{2}$ in the temperature range
between 1.8K and 300K in a double logarithmic plot.
The straight line represents a \mbox{$T^3$ fit} to the
data
at low temperatures.}}
\label{fig4}
\end{center}
\end{figure}
No kink-anomaly of $\kappa$ could be detected at the superconducting phase transition
as reported for the non-magnetic borocarbides \cite{Sera,Boaknin}.
The additionally observed peak below $T_c$ in $\kappa (T)$ of these compounds
should not be regarded as a generic intrinsic feature of
clean superconductors since
investigations of
niobium samples exhibit such a peak only for medium clean
samples \mbox{(RRR$\approx 3000)$} while it
 disappears, if the sample quality is further increased to
\mbox{RRR$\approx 33000$
\cite{Gladun}.}
Furthermore, the reduction of
a possible peak is in agreement with the high phonon velocities
in \mbox{MgB$_2$ \cite{Boaknin,Ravindran}.}

According to
the decrease of the electronic thermal conductivity below $T_c$ a significant
change in the slope of $\kappa (T)$ is found.
Below 7K, where the influence of the electrons is negligible,
the expected \mbox{$T^3$ law} for $\kappa (T)$ 
dominated by phonons at low temperatures 
is observed.
With the measured value of \mbox{$\kappa /T^3=2.9\times 10^{-3}$W/K$^4$m} a
mean free path $l$ of the phonons at low temperatures can be calculated:
\begin{equation}
\l=3\frac{\kappa (T)}{c(T)}\frac{1}{v}V_M,
\label{grain}
\end{equation}
where $c$ is the specific heat, $v$ the acoustic sound velocity, and $V_M$ the
molar volume.
For the lattice contribution of the specific heat $c(T)$
a coefficient
\mbox{$\beta =c/T^3=1.04\times 10^{-5}$J/K$^4$mol} was reported \cite{Kremer}.
Using the calculated values for the sound velocity of \mbox{$v=10600$m/s} for a
longitudinal wave \cite{Ravindran} and a
molar volume of \mbox{$V_M=17.5$cm$^3$/mol}
as derived from
the unit cell volume of \mbox{$V$=29.02 \AA$^3$ \cite{Jorgensen},}
Eq.\ \ref{grain} yields \mbox{$l=1.4\mu $m.} This
length can be interpreted as an averaged grain size and is in agreement with
the results of optical
investigations of the pellets.

The reduced Lorenz number
\begin{equation}
\frac{L(T)}{L_0}=\frac{\kappa (T)\rho (T)}{L_0 T},
\label{Lorenz}
\end{equation}
where $L_0=$2.44$\times 10^{-8}$W$\Omega$/K$^2$ is the Sommerfeld value,
was
derived from the
measured values of
$\kappa (T)$ and $\rho (T)$. The results are shown in \mbox{Fig. 5.} The error
is about \mbox{30\%.}

\begin{figure}[h]
\begin{center}
\epsfig{file=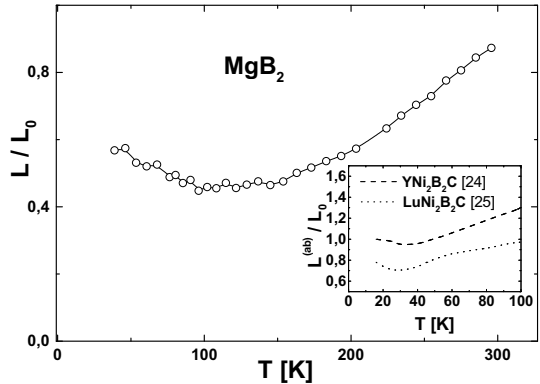,width=8.0cm}
\caption{\footnotesize{Reduced Lorenz number $L/L_0$ of MgB$_{2}$ in the temperature range
between 40K and 300K. The inset shows the reduced Lorenz number of
YNi$_2$B$_2$C and LuNi$_2$B$_2$C single crystals
of the in-plane heat and charge transport properties
taken from Refs. \cite{Sera,Boaknin}. }}
\label{fig5}
\end{center}
\end{figure}
At least for temperatures below 250K the reduced Lorenz number is significantly
smaller than the expected value \mbox{of 1.} Similar results
shown in the inset of\mbox{ Fig. 5} were found for
typical non-magnetic borocarbides
\cite{Sera,Boaknin} and interpreted as the influence of inelastic scattering on the
electronic thermal conductivity. The very small values
of the Lorenz number
confirm the dominating role of
the electronic contribution to the heat transport.

Noteworthy, the shape of the Lorenz plots for MgB$_{2}$ and YNi$_2$B$_2$C /
LuNi$_2$B$_2$C
is very similar. Furthermore, the minima in the reduced Lorenz number occur at temperatures
of about 120K and 35K for MgB$_{2}$ and YNi$_2$B$_2$C / LuNi$_2$B$_2$C,
respectively. These values
correspond to \mbox{$2-3\ T_c$} for that compounds.

Different from these results, the published data of electrical resistivity and thermal
conductivity of Bauer \mbox{{\it et al.\/} \cite{Michor}} yield reduced Lorenz numbers
of up to about 2.5 at room temperature
mainly
caused by the high resistivity values.
Their reported $\rho(300\mathrm{K})\approx70\mu \Omega$cm is
about twice as high as the value found here.
Further investigations are required to clarify these
differences in $L$. However, their reduced Lorenz number
shows a similar behaviour
with a minimum at about 120K.

To summarize, in addition to a number of
well-known
similarities in the
superconducting properties of
MgB$_{2}$ and YNi$_2$B$_2$C, in the present work also
similarities related to the normal state transport properties
have been found in
the temperature dependences of the
resistivity and of
the reduced Lorenz number.
However, the positive sign of the thermopower
as observed in the measurement and derived from bandstructure calculations
indicates
a dominant hole contribution in MgB$_2$. The magnitude of the thermopower
might indicate an intermediate to strong electron-phonon coupling scenario.
The thermal conductivity strongly deviates from the behaviour expected for
clean samples. The averaged grain size of the sample inferred from this
measurement is \mbox{about $1.4\mu $m.}

\vspace{15mm}

After completion \cite{version0} of the present work, we have learnt about a preprint by
 Putti \mbox{{\it et al.\/} \cite{Putti}}. 
 Their resistivity data differ from ours by 
 a significantly higher residual resistivity and a smaller RRR-value.
 The data have been described by a generalized Bloch-Gr\"uneisen
formula. The raw thermopower data $S(T)$ looks very similar to ours. 
However, the interpretation is different.
In the narrow interval \mbox{$45$K$<T<90$K} $S(T)$ 
was fitted to a linear term $S_{el}$ and a {\it positive} cubic one 
({\it i. e.\/} the low-temperature approximation for the phonon drag contribution)
which would point to predominant N(normal) scattering processes at low temperature.
The high temperature region was not quantitatively analyzed. 
Our data can be analyzed by those terms only in the range \mbox{$55$K$<T<90$K} 
with clear deviations above $90$K and also {\it below} $55$K.   
Anyhow, this should be compared 
with our fit (Eq.\ (2)) in a broader interval \mbox{$70$K$<T<270$K}. It contains a 
negative high-temperature approximation for the phonon drag contribution 
which can be interpreted in terms of  U(Umklapp)-processes \cite{Blatt} and the 
relevance of soft modes clearly below the Debye energy. In this context 
the observation of low energy peaks at about 16 meV and 24 meV in recent neutron scattering 
\cite{Muranaka0} and 17 meV in Raman measurements \cite{Lampakis} is of interest. 
Naturally, due to these different adopted approximations for the phonon drag terms 
with opposite signs, 
different dressed linear electronic diffusion terms \mbox{$S_{el}/T=B=0.0176 \mu$V/K$^2$} 
and \mbox{$B=0.031 \mu$V/K$^2$} 
have been derived. Since
the band structure result for the bare $S_{el}$-term, calculated 
in the same 
approximation for the scattering rate as we did, seems to coincide with our 
result, different renormalizations due to many-body effects would be expected.

Furthermore, Muranaka \mbox{{\it et al.\/} \cite{Muranaka}} reported a
saturation of the thermopower near room temperature 
at a relatively low level of only 4$\mu$V/K
(compared with about 8 $\mu$V/K in our work or in Ref.\ \cite{Putti})
and a reduced Lorenz number
with a similar shape and a magnitude in between that of 
Bauer \mbox{{\it et al.\/} \cite{Michor}}
and that of the present work. 
All these different features mentioned above require further
investigations especially with respect to the sample quality.


\vspace{25mm}

This work has been supported by the SFB 463,
 the DFG, the DAAD (individual grant, H.R.),
and the ONR Grant No. N00017-97-1-0956.
We acknowledge valuable discussions with S.V.\ Shulga, I.\ Mertig,
and H.\ Eschrig.

\end{document}